\documentclass[preprint,preprintnumbers]{revtex4}

\usepackage{graphicx}

\begin{document}



\title{A scaling approach for  interacting quantum wires - A possible explanation for the 0.7 anomalous conductance}

\author{ D. Schmeltzer$^{1}$, A. Kuklov$^{2}$ and M.Malard$^{3}$}

\address{$^{1}$  Department of Physics,  City College  of the CUNY, $^{2}$ Physics Department at the College of Staten Island of the CUNY, $^{3}$ International  Center for Condensed Matter Physics Brasilia/Brazil}

\begin{abstract}

We consider a weakly interacting finite wire    with short and long range   interactions. The  long range interactions  enhance    the  $4k_{F}$  scattering and renormalize the wire to   a strongly interacting limit.
For large screening lengths,  the renormalized  charge stiffness Luttinger parameter $K_{eff.}$ decreases to $K_{eff.}< \frac{1}{2}$, giving rise to a Wigner crystal at $T=0$ with  an anomalous conductance at finite temperatures.
 For short  screening lengths, the renormalized Luttinger   parameter $K_{eff.}$ is  restricted  to  $\frac{1}{2}\leq K_{eff.}\leq 1$. As a result,   at  temperatures larger than the magnetic exchange energy we find     an  interacting metal   which for   $K_{eff.}\approx \frac{1}{2}$  is equivalent to the   Hubbard $U\rightarrow\infty$ model, with the  anomalous conductance  $G\approx\frac{e^2}{h}$ .

\end{abstract}


\maketitle

\vspace{0.1 in}

\textbf{1. Introduction}

\vspace{0.1 in}

The  anomalous conductance  $G\approx 0.7\times(2e^{2}/h)$ discovered   by Pepper  et al.
 \cite{Pepper} and further investigated  by \cite{Pepperoni,Cronennwett,Picciotto,Bird,Komijani} is one of the major   unexplained effects  in quantum wires.   Several   theories have  been proposed: Phenomenological theories \cite{Bruus,Reilly}, Kondo effect  \cite{Cronennwett}, spin polarization \cite{Meir,Kaveh}, formation of bound states \cite{Bird}, unrestricted Hartree-Fock calculations for point contacts \cite{Sushkov}, Wigner crystal \cite{Matveev}, ferromagnetic spin coupling \cite{Han}, ferromagnetic zigzag structures  \cite{Meyer},   the  spin incoherent Luttinger liquid 
\cite{Fiete,Siljuasen,Cheianov}, and the formation of  a quasi-localized state \cite{Rejec}, however no consensus   has been reached.

For noninteracting spin unpolarized electrons, the conductance of  narrow ballistic quantum wires connected to two  (large) reservoirs is quantized in units of $2e^2/h$. 
An early suggestion was that the  electron-electron interaction should modify the conductance for a Luttinger liquid \cite{Haldane},\cite{Kane}  as $K(2 e^2/h)$  where $K$ is interaction-dependent.  Using the method of Bosonization for weakly  interacting fermions, it has been shown that by taking first the  frequency limit $\omega\rightarrow 0$ before the momentum limit $q\rightarrow 0$, the non-interacting leads modify the metallic conductance to  the limit  $G = \frac{2e^{2}}{h}$  \cite{Safi,Maslov,David} .
 
 $GaAs/AlGaAs$ in the lowest populated conduction band is  a  weakly interacting metal characterized by the Luttinger liquid  charge  $K\leq 1$ and spin $K_{s}\approx1$ parameters. The presence of  the unscreened, long range Coulomb interaction  (in a one dimensional wire)  alters this picture. From our Renormalization Group study \cite{Mariana} we find that the Coulomb long range interaction enhances  the  weak $4k_{F}$ scattering channel and  decreases  the Luttinger charge parameter to  $K_{eff.}<<1$ .
 As a function of the screening length, we can have either a  strongly interacting metal (similar to the Hubbard $U=\infty$ model)  or  an insulating  Wigner crystal.  
In order to capture both  phases, we investigate a microscopic model of a wire of length $L$ with a lattice constant $a$ in the presence of a  weak  scattering  potential. 
The  lattice constant $a$  is   much smaller than the transverse  width  $d$ of the wire,   which controls   the  low energy excitations in the lowest band. The effective model in the lowest band  is  given by the renormalized   lattice model. This is   achieved by a Real Space Renormalization Group procedure, which replaces  the discrete lattice model  (lattice constant $a$)  by  the new lattice  constant $d=Integer\times a$,   and   renormalized   interaction coupling constants.    
  The  effective model at the length scale $d$  will be a function of  the microscopic Fermi momentum $k_{F}$ (defined by the electronic density) and    the effective umklapp momentum $G^{(d)}$   obeying the relation  $G^{(d)} \cdot d=G^{(a)}\cdot a$, where  $G^{(a)}$ is  the  microscopic umklapp vector.  At the length scale $d$, the long range interaction is separated into two parts: the large momentum transfer, included  in the effective short range Hubbard interaction, and the $forward$
 Coulomb interaction. The forward Coulomb interaction   gives rise to $4k_{F}=G^{(Wigner)}=\frac{2\pi}{r_{e-e}}$   oscillations \cite{Shultz} where $r_{e-e}$ is the inter-particle distance.  At T=0,  a  Wigner crystal ground state   with a charge gap  $\Delta$ is formed if the effective charge parameter obeys  $K_{eff.}<\frac{1}{2}$.  
  At finite temperatures,   comparable to the charge gap $\Delta$, the conductance  is given by       $G\approx  \frac{e^{2}}{h}$. 
For short  screening lengths,  the  interacting parameter  $K_{eff.}$ is restricted to $\frac{1}{2}\leq K_{eff.}<1$ .  As a result, we find that at finite temperatures which are larger than the magnetic exchange  energy, 
 the limit $K_{eff.}\approx\frac{1}{2}$ is equivalent to the Hubbard $U=\infty$ model. Therefore,  at finite temperatures  we find the conductance  is given by  $G\approx \frac{e^2}{h}$.

The plan of this paper is as follow: In chapter $2$, we present the interacting Fermion model. The renormalization effects for  the finite wire  will  be investigated in chapter $3$ using  the  zero mode  formulation  \cite{David,Avadh,Mathieu}. In chapter $4$,  we present the Fermion-Boson representation  for the interacting wire.      
In chapter $5$, we present the Renormalization Group (R.G.)  analysis   and show that the    Renormalization effects of the   effective charge interacting parameter $K_{eff.}$  are controlled by the electronic density and screening length.
In chapter $6$,  we use the renormalized  interaction parameters to compute the effective zero mode Hamiltonian  at finite temperatures for $\frac{1}{2}\leq K_{eff.}<1$.   Chapter $7$ is dedicated to the computation of the conductance   at finite temperatures for $\frac{1}{2}\leq K_{eff.}<1$. In chapter $8$, we consider the case  $\frac{1}{2}\approx K_{eff.}$ and show that the model is equivalent  to the incoherent Luttinger liquid which emerges at finite temperatures for the Hubbard $U\rightarrow\infty$ model.
In chapter $9$, we consider the case   $K_{eff.}< \frac{1}{2}$  which  at zero temperature   gives rise to a Wigner crystal   with a charge density wave gap $\Delta $.  
In chapter $10$, we present our numerical results  using the experimental parameters given by \cite{Picciotto}. In chapter $11$, we examine the effect of the Zeeman interaction.  Chapter  $12$ is devoted to  conclusions.  Appendix $A$  deals with the thermodynamics of the zero modes,  and in Appendix  $B$  we present the calculation of the self energy  for a  wire  of length $L\approx 10^{-6}$  at  temperatures $T\approx 1$ Kelvin.

\vspace{0.1 in} 

\textbf{2. The  model}  
 
\vspace{0.1 in}

We  consider an interacting   wire at  low electronic densities that has  a finite width $d$. The  geometric parameters   in the quantum wire experiments are: the gate  screening  length    $\xi= 10^{-7} m$, the wire length $L\approx 10^{-6} m$, the  width   $d\approx  \frac{L}{100}$,  the  two dimensional carrier density  $n_{s}\approx 2.5\times 10^{11} cm^{-2}$ and the electronic lattice spacing   $a\approx  10^{-10} m$. Due to the width  $d$,   the single particle excitations are characterized by  a set of  electronic bands with the  transverse quantization energies $\frac{\hbar^2}{2m*}\frac{(r^2)\pi^2}{d^2}$,   $r=1,2,3...$. The gate voltage is such that, at the   temperatures considered in the experiment,  only the $lowest$ $band$ is  populated. In order to describe the low energy physics, we  project the microscopic Hamiltonian  $H^{a}$  (the microscopic model defined at the lattice scale $a$)  into the lowest band.  As  a result, the   effective one dimensional Hamiltonian $H$ with the  energy cut-off characterized by the transversal energy separation   $\frac{\hbar^2}{2m*}\frac{\pi^2}{d^2}$ and  the momentum cut-off $\Lambda=\frac{2\pi}{d}$   
preserve the original form  of the microscopic Hamiltonian  $H^{a}$: 
\begin{eqnarray}
H&=&-t_{d}\sum _{n}\sum_{\sigma=\uparrow,\downarrow}(\psi^{+}_{\sigma}((n+1)d)\psi_{\sigma}(nd)+h.c.)-\epsilon_{F}\sum_{n}\sum_{\sigma=\uparrow,\downarrow}\psi^{+}_{\sigma}(nd)\psi_{\sigma}(nd) \nonumber\\&& + \hat{U}\sum_{n}\psi^{+}_{\downarrow}(nd)\psi_{\downarrow}(nd)\psi^{+}_{\uparrow}(nd)\psi_{\uparrow}(nd) \nonumber\\&&+\sum_{n}\sum_{n'\neq n}\sum_{\sigma=\uparrow,\downarrow}\sum_{\sigma'=\uparrow,\downarrow}\psi^{+}_{\sigma}(nd)\psi_{\sigma}(nd)V^{(c)}(|nd-n'd|) \psi^{+}_{\sigma'}(n'd)\psi_{\sigma'}(n'd)
\label{eqnarray}
\end{eqnarray}
where $t_{d}\approx\frac{\hbar^2}{2m* d^2}$ is the effective hopping at  the length scale $d$, $\hat{U}$ is the projected   repulsive  Hubbard interaction,   which also contains  the effect of the Coulomb interaction obtained by projecting out states  with a lattice spacing in the interval $a-d$. $ V^{(c)}(|nd-n'd|)=\frac{e^2}{\sqrt {(n-n')^{2}d^2+d^2}}-\frac{e^2}{\sqrt{ (n-n')^{2}d^2+\xi^2}}$    is the  effective long range Coulomb interaction defined for distances $x>d$ and $\xi$ is the screening length.  
The  Hubbard model is characterized by the particles - holes charge and spin  excitations:   $K\leq1 $ (charge)  and $K_{s}\geq 1$  ( spin ).  We consider the  situation   where  the Fermi momentum   $k_{F}(V_{G})$   satisfies the condition  $4k_{F}(V_{G})\neq G^{d}\equiv\frac{2\pi}{d}$, $r_{e-e}>d$,  suggesting  that the  umklapp interaction is negligible.
According to    \cite{Picciotto}, the density is expressed   in terms of  the external gate voltage $V_{G}$: $k_{F}(V_{G})=\frac{\pi}{2}n_{e}(V_{G})=\frac{C_{a}}{e}(V_{G} -V^{th})$, where    $V^{th}$ is the gate voltage at which the wire is pinched  off and  $\frac{C_{a}}{e}$ is the effective capacitance.
 The   results  reported in \cite{Picciotto} show that the  conductance decreases with the lowering of  the gate voltage,  suggesting that the umklapp interaction  is  significantly enhanced.
  The low energy properties of the model will be investigated using  a combined method of Bosonization and  R.G. theory. At  finite temperatures, the exact description of the  electron excitations requires the inclusion of the    $zero-modes$ operators.

\vspace{0.1 in}

\textbf{3. The representation of the electron operator for  a wire of length $L$}

\vspace{0.1 in}

The   electron  is represented as a product of two operators, a $Bosonic$ one (this is the standard Bosonic representation for spin-charge  excitations) and  a $Fermionic$-$zero$ $mode $ operator, which carries  the electron number (electrons with spin up or spin down that  are added or removed from the Fermi surface).       
The  electron operator   $\psi^{+}_{\sigma}(x)$ is restricted by the momentum with   a   momentum  cut-off  $[\Lambda,-\Lambda]$ around the Fermi surface   and  is given in terms of the  right $R_{\sigma}(x)$ and left $L_{\sigma}(x)$  components: 
 $\psi_{\sigma}(x)=e^{i k_{F}x}R_{\sigma}(x)+ e^{-i k_{F}x}L_{\sigma}(x)$ with the  Fermi momentum $k_{F}= k_{F}(V_{G})$.
  We replace the right  (left) mover fermion  by a product of a fermion   operator  $F_{R,\sigma}(x)$ ($F_{L,\sigma}(x)$) and the boson one  $e^{i\sqrt{4\pi}\Theta_{R,\sigma}(x)}$ ($e^{i\sqrt{4\pi}\Theta_{L,\sigma}(x)}$): 
 \begin{equation}
R_{\sigma}(x)={\sqrt{\frac{\Lambda}{2\pi}}} e^{i\alpha_{R,\sigma}}e^{i(2\pi/L)(\hat{N}_{R,\sigma}-1/2)x}e^{i\sqrt{4\pi}\Theta_{R,\sigma}(x)}\equiv F_{R,\sigma}(x)e^{i\sqrt{4\pi}\Theta_{R,\sigma}(x)}
\label{right}
\end{equation}
\begin{equation}
L_{\sigma}(x)={\sqrt{\frac{\Lambda}{2\pi}}} e^{-i\alpha_{L,\sigma}}e^{-i(2\pi/L)(\hat{N}_{L,\sigma}-1/2)x}e^{-i\sqrt{4\pi}\Theta_{L,\sigma}(x)}\equiv F_{L,\sigma}(x)e^{-i\sqrt{4\pi}\Theta_{L,\sigma}(x)}
\label{left}
\end{equation}
where $e^{i\sqrt{4\pi}\Theta_{R,\sigma}(x)}$ and  $e^{-i\sqrt{4\pi}\Theta_{L,\sigma}(x)}$ are the standard Bosonization  formulas used in the literature.  $F_{R,\sigma}(x)$ and $F_{L,\sigma}(x)$ are the zero mode fermion operators that  can change the number of particles and are crucial for enforcing the Fermi  statistics. This operators are defined with respect to the  non - interacting ground state $|F>$ (the  Fermi see characterized by the Fermi momentum).
The electronic Hilbert space  excitations above the Fermi see  are given by  the states  \cite{David,Avadh}:  $|N_{R,\sigma};m_{q}>\otimes|N_{L,\sigma};m'_{q}>$ where  $m_{q}\geq 0$ ,  $m'_{q}\geq 0$ are integers which specify the number of Bosonic quanta  (particles -holes excitations) with a momentum $q=\frac{2\pi}{L}n_{q}>0$.     $\hat{N}_{R,\sigma}$ , $\hat{N}_{L,\sigma}$ represent the change  of the total number of electrons  in   the right and left  ground states. The formal proof that  relates the  electron operator   to the zero mode operators  is given  by  the   Jacoby  identity \cite{Mathieu}.  
 The Bosonic - particle hole  excitations \cite{David,Avadh} are given by: $\Theta_{R,\sigma}(x)=\Theta_{\sigma}(x)-\Phi_{\sigma}(x)$,  $\Theta_{L,\sigma}(x)  =\Theta_{\sigma}(x)+\Phi_{\sigma}(x)$. The  zero mode Fermion excitations are given in terms of   the zero mode coordinates $\alpha_{R,\sigma}$,$\alpha_{L,\sigma}$  and  their canonical conjugate fermion number operators  $\hat{N}_{R,\sigma}$ , $\hat{N}_{L,\sigma}$ ( $\sigma=\uparrow,\downarrow$).
 The physics of the zero modes is described in terms of the  $charge$ operator $\hat{Q}_{c}=\sum_{\sigma=\downarrow,\uparrow} [\hat{N}_{R,\sigma}+\hat{N}_{L,\sigma}]$ and the  $magnetization$ operator $\hat{Q}_{s}= [(\hat{N}_{R,\sigma=\uparrow}+\hat{N}_{L,\sigma=\uparrow})-(\hat{N}_{R,\sigma=\downarrow}+\hat{N}_{L,\sigma=\downarrow}]$.
The canonical conjugate variables to the $charge$ and $magnetization$  are given by the $charge$ $coordinate$    $\hat{\alpha}=\sum_{\sigma=\downarrow,\uparrow}[ \alpha_{R,\sigma}+\alpha_{L,\sigma}]$ and the $magnetization$ $coordinate$  $\hat{\alpha}_{s}= [( \alpha_{R,\sigma=\uparrow}+\alpha_{L,\sigma=\uparrow})-( \alpha_{R,\sigma=\downarrow}+\alpha_{L,\sigma=\downarrow})]$.
 The zero modes obey the commutation rules : $[\alpha_{R,\sigma},\hat{N}_{R,\sigma'}]=i\delta_{\sigma,\sigma'}$, $[-\alpha_{L,\sigma},\hat{N}_{L,\sigma'}]=i\delta_{\sigma,\sigma'}$ and $[\alpha_{R,\sigma},\hat{N}_{L,\sigma'}]= [\alpha_{L,\sigma},\hat{N}_{R,\sigma'}]=0$.
 
\vspace{0.1 in}

\textbf{4. The model  Hamiltonian  in the Boson-Fermion  representation}

\vspace{0.1 in}

The  Bethe Ansatz  formulation \cite{Kawakami}  and  eqs. $(2-3)$ allows  us to  map   eq. $(1)$ into a charge and spin interacting  model. The  mapping  is a function of the Hubbard interaction  strength $U\equiv\frac{\hat{U}}{t_{d}}$ and  the electronic density $n_{e}$. The  Hamiltonian is controlled by  the charge  parameter  $K=K(U,n_{e})$ , spin parameter $K_{s}= K(U,n_{e})$, umklapp interaction   $g=g(U,n_{e})=\hat{g}\Lambda^2$, spin  backward scattering parameter  $g_{s}=g(U,n_{e})=\hat{g}_{s}\Lambda^2$,  Fermi  velocity  $v_{F}$, charge density  wave  velocity  $v=v(U,n_{e})$  and  the spin  density wave velocity  $v_{s}=v_{s}(U,n_{e})$. 
The Hamiltonian in eq.$(1)$ is replaced by: $H=H^{n\neq0}_{c}+H^{n\neq0}_{s}+H^{(n=0)}$.    
The first two Hamiltonians $H^{n\neq 0}_{c}+H^{n\neq 0}_{s}$ represent the particle hole excitations  and $H^{(n=0)}$ represents the zero modes.
 The  charge excitations  $H^{n\neq 0}_{c} (\Theta,\Phi;\hat{\alpha},\hat{Q_{c}})$ are given   in terms of the Bosonic fields $\Theta=\frac{\Theta_{\uparrow}+\Theta_{\downarrow}}{\sqrt{2}}$, $\Phi=\frac{\Phi_{\uparrow}+\Phi_{\downarrow}}{\sqrt{2}}$  and the zero mode Fermionic fields $\hat{\alpha}$,$\hat{Q_{c}}$: 
\begin{eqnarray}
H^{n\neq 0}_{c}(\Theta,\Phi;\hat{\alpha},\hat{Q}_{c})&=&v\hbar[\int_{-L/2}^{L/2}\,dx[\frac{K}{2}(\partial_{x}\Phi(x))^{2}+\frac{1}{2K}(\partial_{x}\Theta(x))^{2}\nonumber\\&  & -g\cos[\sqrt{8\pi }\Theta(x)+\hat{\alpha}+(4k_{F}(V_{G})+\frac{2\pi}{L}\hat{Q_{c}})x]]] \nonumber\\&  &+\frac{e^{2}}{\pi\kappa_{0}}\int_{-L/2}^{L/2}\int_{-L/2}^{L/2}\,dx\,dx'\partial_{x}\Theta(x)[\frac{e^2}{\sqrt {(x-x')^{2}+d^2}}-\frac{e^2}{\sqrt{ (x-x')^{2}+\xi^2}}]\partial_{x'}\Theta(x')\nonumber\\&
\end{eqnarray}
 where     $v$  is the charge velocity  $v K=v_{F}=\frac{\hbar K_{F}(V_{G})}{m*}$  and $K=K(U,n_{e})$ is the  charge interaction parameter. 
The last term  in equation $(4)$ represents the  $forward$ part of the long range interaction given in eq.$(1)$ with the screening length $\xi$.  The long range interactions  is controlled  by the coupling constant     $\gamma=\frac{e^2}{\hbar c}\cdot\frac{1}{\kappa_{0} }$, where    $c$  is the  light velocity and $\kappa_{0} = 13.18$  is the dielectric constant for $GaAs$.
The strength of the umklapp interaction $g$ is determined by two parts: the  short range Hubbard $U $  repulsive interaction and the large momentum transfer of the Coulomb interaction  obtained after the projection.
The spin density  wave excitations are given by  the Hamiltonian $H^{n\neq 0}_{s} (\Theta_{s},\Phi_{s};\hat{\alpha}_{s},\hat{Q}_{s})$  with 
the spin density  wave operators:  $\Theta_{s}=\frac{\Theta_{\uparrow}-\Theta_{\downarrow}}{\sqrt{2}}$, $\Phi_{s}=\frac{\Phi_{\uparrow}-\Phi_{\downarrow}}{\sqrt{2}}$    
\begin{eqnarray}
&&H^{n\neq 0}_{s}(\Theta_{s},\Phi_{s};\hat{\alpha}_{s},\hat{Q}_{s})=v_{s}\hbar[\int_{-L/2}^{L/2}\,dx[\frac{K_{s}}{2}(\partial_{x}\Phi_{s}(x))^{2}+\frac{1}{2K_{s}}(\partial_{x}\Theta_{s}(x))^{2}\nonumber\\&& +g_{s}\cos(\sqrt{8\pi }\Theta_{s}(x)+\hat{\alpha}_{s}+\frac{2\pi}{L}\hat{Q_{s}}x)]]
\end{eqnarray}
where $ v_{s}= \frac{v_{F}}{K_{s}}<v$ is the spin wave velocity, $K_{s}>1$ is the spin stiffness and  $g_{s}$ is the   spin density wave coupling constant.
Next we present  the zero mode Hamiltonian  $H^{(n=0)}$ using the $normal$ $order$ notation:
 $:H^{(n=0)}:\equiv H^{(n=0)}-<F|H^{(n=0)}|F>$ where $|F>$ is  the unperturbed Fermi surface  at zero temperature  given in terms of shifted operators, 
  $N_{R,\sigma}=\hat{N}_{R,\sigma}+<F|N_{R,\sigma}|F>$;
  $N_{L,\sigma}=\hat{N}_{L,\sigma}+<F|N_{L,\sigma}|F>$ (see Appendix $A$). The  normal order, zero mode Hamiltonian takes the form: $:H^{(n=0)}:=  :H_{0}^{(n=0)}:+ :H_{int}^{(n=0)}:$.     
\begin{equation}
:H_{0}^{(n=0)}:=\frac{h v_{F}}{2L}[\hat{N}^{2}_{R,\sigma=\uparrow}+\hat{N}^{2}_{L,\sigma=\uparrow}+\hat{N}^{2}_{R,\sigma=\downarrow}+\hat{N}^{2}_{L,\sigma=\downarrow}]
\label{zerom}
\end{equation}
\begin{eqnarray}
&&:H_{int}^{(n=0)}:= u^{(c)}(L)[(\hat{N}_{R,\sigma=\uparrow}+\hat{N} _{L,\sigma=\uparrow})+(\hat{N}_{R,\sigma=\downarrow}+\hat{N} _{L,\sigma=\downarrow})]^2\nonumber\\&&
-u^{(s)}(L)[(\hat{N}_{R,\sigma=\uparrow}+\hat{N} _{L,\sigma=\uparrow})^{2}-(\hat{N}_{R,\sigma=\downarrow}+\hat{N} _{L,\sigma=\downarrow})^2]\nonumber\\&&+ \frac{e^2}{\kappa_{0} }\frac{1}{2L}F(\frac{L}{d },\frac{\xi}{d })[(\hat{N}_{R,\sigma=\uparrow}+\hat{N} _{L,\sigma=\uparrow})+(\hat{N}_{R,\sigma=\uparrow}+\hat{N} _{L,\sigma=\uparrow})]^{2}
\end{eqnarray}
The zero mode coupling constants  obtained from eq.$(1)$   are given by  the renormalized  charge backward interaction  $u^{(c)}(L)=\frac{h v_{F}}{2L}(\frac{1-K^{2}}{K^{2}})$  and the backward  spin  interaction   $u^{(s)}(L)=\frac{h v_{F}}{2L}(\frac{1-K^{2}_{s}}{K^{2}_{s}})$. At   zero temperature and $L\rightarrow \infty$,  $K_{s}$ flows to 1  and the backward  interaction   $u^{(s)}(L)$  vanishes. The function  $F(\frac{L}{d },\frac{\xi}{d })=log[\frac{\sqrt{[1+(\frac{d }{L})^2]}+1}{\sqrt{[1+(\frac{d }{L})^2]}-1}]-log[\frac{\sqrt{[1+(\frac{\xi }{L})^2]}+1}{\sqrt{[1+(\frac{\xi }{L})^2]}-1}]$ is the    Fourier transform  of the long range   screened  potential. At finite temperatures, the Fermi energy is shifted by  $\delta\mu_{0}(T)$  which modifies the zero mode Hamiltonian: $\delta H^{(n=0)}=\delta\mu_{0}(T)[(\hat{N}_{R,\sigma=\uparrow}+\hat{N} _{L,\sigma=\uparrow})+(\hat{N}_{R,\sigma=\uparrow}+\hat{N} _{L,\sigma=\uparrow})]$. 

\vspace{0.1 in}

\textbf{5. The Renormalization Group equations}

\vspace{0.1 in}

 One of us  \cite{ScRG} has developed an R.G. method which is applicable for  the Hamiltonian representation. This method has been used  \cite{Mariana} to derive the R.G. equations for the  $unbiased$  $Sine$-$Gordon$ in the presence of long range interactions  controlled  by the coupling constant     $\gamma=\frac{e^2}{\hbar c}\cdot\frac{1}{\kappa_{0} }$.
\begin{eqnarray}
&&H^{n\neq 0}_{c}(\Theta,\Phi)=v\hbar[\int_{-L/2}^{L/2}\,dx[\frac{K}{2}(\partial_{x}\Phi(x))^{2}+\frac{1}{2K}(\partial_{x}\Theta(x))^{2} -g\cos(\sqrt{ 2n 8\pi }\Theta(x))]] \nonumber\\&  &+\frac{e^{2}}{\pi\kappa_{0}}\int_{-L/2}^{L/2}\int_{-L/2}^{L/2}\,dx\,dx'\partial_{x}\Theta(x)[\frac{e^2}{\sqrt {(x-x')^{2}+d^2}}]\partial_{x'}\Theta(x')\nonumber\\&
\end{eqnarray}
 In the absence of the Coulomb interaction  the   model is  equivalent to the classical two dimensional Sine-Gordon model.
According to  \cite{Koster,Berez} the  model is gaped   for $K<\frac{1}{2n}$, $n=1,2...$. The  long-range interaction modifies  the results and  drives the model  to a gaped phase for any value of $K$.
We have  extended   the R.G. calculations   for the  $biased$  Sine Gordon model
 $g\cos[\sqrt{8\pi }\Theta(x)+\hat{\alpha}+(4k_{F}(V_{G})+\frac{2\pi}{L}\hat{Q_{c}})x]$ given in eq.$(4)$. 
We obtain the $new$  R.G.  equations  as function of the $bias$ and  the $screening$ length  $\xi$  for the differential  momentum shell  $dl=-\frac{d\Lambda}{\Lambda}$.  
\begin{equation}
(4k_{F}(V_{G})+\frac{2\pi}{L}\hat{Q_{c}})\rightarrow (4k_{F}(V_{G})+\frac{2\pi}{L}\hat{Q_{c}})e^{l}
\label{bias}
\end{equation}
\begin{equation}
\frac{d\hat{g}_{R}(l)}{dl}= 
   2 \hat{g}_{R}(l) (1 - 
     \frac{ K_{R}(l)}{\sqrt{(1 + \gamma(\frac{ c}{v_{R}(l)})M_{R}(l))}} - \frac{K^{2}_{R}(l)         \hat{g}^{2}_{R}(l)  }{4 (1 + \gamma(\frac{ c}{v_{R}(l)})M_{R}(l))})
\label{u}          
\end{equation} 
\begin{equation}         
\frac{d K_{R}(l)}{d l}=-  \frac{(K^{3}_{R}(l) \hat{g}^{2}_{R}(l)  }{8(1 + \gamma\frac{ c}{v_{R}(l)}  M_{R}(l))} 
\label{k}
\end{equation}
\begin{equation}
\frac{d v_{R}(l)}{dl}=
 \frac{v_{R}(l) K_{R}(l)^2 \hat{g}^{2}_{R}(l)}{4 (1 + \gamma\frac{ c}{v_{R}(l)}  M_{R}(l))} 
\label{v}
\end{equation}
where $M_{R}(l)$ is the difference of two  Bessel functions $ K_{0}(x)$:
$M_{R}(l)= 2(K_{0}[e^{-l}] - K_{0}[\frac{\xi}{d}\cdot e^{-l}])$
The solution of the R.G. equations depends on the  initial values of the interaction parameters  $\hat{g}_{R}(l=0)$ , $K_{R}(l=0)$  and  the ratio  $\frac{\xi}{d}$.
We will  study  the case where  $4k_{F}(V_{G})\leq\frac{\pi}{d}$. In order to compute the scaling functions, we need to determine  the relation between the logarithmic scale   $l$ and the voltage  $V_{G}$.  
Based on the experimental observation  \cite{Picciotto} we have  a perfect conductance for a particular gate voltage   $V^{(0)}_{G}$ for which the  umklapp interaction is negligible. This will happen if        $4k_{F}(V^{(0)}_{G})$  corresponds to the momentum  $\frac{\pi}{d}$.  For this case we find  an oscillating behavior for the Sine Gordon term:    $g\cos[\sqrt{8\pi }\Theta(x)+\hat{\alpha}+(4k_{F}(V^{(0)}_{G})+\frac{2\pi}{L}\hat{Q_{c}})x]= g (-1)^{n} \cos[\sqrt{8\pi }\Theta(x)+\hat{\alpha}+(\frac{2\pi}{L}\hat{Q_{c}})x]$ and can ignore the umklapp contribution.  For lower  gate voltages   $V_{G}< V^{(0)}_{G}$  the situation is different.  Following   \cite{Ping} we do not neglect the  umklapp interaction for   $V_{G}< V^{(0)}_{G}$, instead we  compute the effective coupling constant  at the  length scale  $l=l(V_{G})$.  This length scale  $l=l(V_{G})$ is determined   by the  equation   $4k_{F}(V_{G}) e^{l(V_{G})}= 4k_{F}(V^{(0)}_{G})$ and is given by    $l(V_{G})= log[\frac{4k_{F}(V^{(0)}_{G})}{4k_{F}(V_{G})}]$.  At this length scale, the   renormalized umklapp  interaction alternates in sign  $g (-1)^{n}$ and can be neglected if   the Sine-Gordon coupling constant is small. 
Using this procedure we  substitute the function  $l(V_{G})$  into the R.G.  equations   and find the   renormalized Luttinger parameter as a function of the gate voltage $V_{G}$.
Since the wire has a finite length  $L$ we   stop  the scaling   when  we reach, the value $l=minimum[l(V_{G}),l_{L}]$ where $l_{L}=log[\frac{L}{d}]$. 
  In the presence of the Coulomb interactions  the Luttinger parameter  $K_{R}(l(V_{G}))$ is replaced  by the effective  
 parameter 
$K_{eff.}(l(V_{G}))$, computed from the R.G. equations $(9-12)$:
\begin{equation}
K_{eff.}(l(V_{G}))= \frac{ K_{R}(l(V_{G}))}{\sqrt{1+\gamma \cdot\frac{c}{v_{R}(l(V_{G}))}\cdot log[(\frac{\xi}{d})^2]}}
\label{eo}
\end{equation}
The  effective interaction parameter $K_{eff.}(l(V_{G}))$  decreases monotonically  with the decrease in the density and exhibit a maximum for densities where $K_{eff.}(l(V_{G})\approx \frac{1}{2}$. The charge density velocity is  enhanced to  $v=\frac{v_{F}}{K_{eff.}(l(V_{G}))}$. 
When the screening ratio approaches $\frac{ \xi}{d}=1$,  the Coulomb  renormalization is absent and  $K_{eff.}(l(V_{G}))=K_{R}(l(V_{G}))$.
In figure $1$ we have plotted $\frac{1}{K^2_{eff.}(l(V_{G}))}$ as a function of the gate voltage.  Following  \cite{Picciotto},  we have  used for the gate voltage  $V^{(0)}_{G}$ the value  $V^{(0)}_{G}=-5.1$ volt. We observe that $\frac{1}{K^2_{eff.}(l(V_{G}))}$  has a minimum for voltages that corresponds to the region where the $0.7$ feature is seen. Since the compressibility   $\kappa$  is proportional to the square of the Luttinger parameter  $\kappa\propto K^2_{eff.}(l(V_{G}))$,  we conclude that a  maximum in the compressibility suggests the  formation of a  gap. (Since the compressibility is proportional to the derivative of the renormalized chemical potential  $\mu_{R}(V_{G},l(V_{G}))$,  $\frac{1}{\kappa}=(n_{e}(V_{G}))^2 \partial _{n_{e}(V_{G})}[\mu_{R}(V_{G},l(V_{G}))]$ we expect also a minimum for the derivative.) 

\vspace{0.1 in}

\textbf{6. The effective Hamiltonian  $\frac{1}{2}\leq K_{eff.}(l(V_{G}))\leq1$ for $L>\xi$}

\vspace{0.1 in}

Using the dependence of the Fermi momentum  $k_{F}(V_{G})< k_{F}(V^{(0)}_{G})$ on the gate voltage $V_{G}<  V^{(0)}_{G}$,  we find that the umklapp interaction and  the Luttinger  parameter are renormalized. Following the analysis from chapter $5$, we find that at the length scale  $l(V_{G})= log[\frac{4k_{F}(V^{(0)}_{G})}{4k_{F}(V_{G})}]$  the renormalized umklapp interaction  is negligible $g(l(V_{G}))\approx 0$ and the renormalized velocity is   $\frac{v_{F}}{K_{eff.}(l(V_{G})}$. When  $L>\xi$, the effective Luttinger parameter is restricted to  $\frac{1}{2}\leq K_{eff.}(l(V_{G})\leq1$.   
The renormalized  Bosonic  Hamiltonian is given by:
\begin{eqnarray}
&&H^{(n\neq0)}_{c,l(V_{G})}[\Theta_{R},\Phi_{R}]\approx \hbar v_{R}(l(V_{G}))[\int_{-L/(2e^{l(V_{G})})}^{L/(2e^{l(V_{G})}) }\,dx[\frac{K_{R}(l(V_{G}))}{2}(\partial_{x}\Phi_{R}(x))^{2}+\frac{1}{2K_{R}(l(V_{G}))}(\partial_{x}\Theta_{R}(x))^{2}]]\nonumber\\&  &+\frac{e^{2}}{\pi\kappa_{0}}\int_{-L/(2e^{l(V_{G})})}^{L/(2e^{l(V_{G})})} \,dx\,dx'\partial_{x}\Theta_{R}(x)[\frac{1}{\sqrt{ (x-x')^{2}+(\frac{d}{e^{l(V_{G})}})^2}}-\frac{1}{\sqrt {(x-x')^{2}+(\frac{\xi}{e^{l(V_{G})}} )^2}}]\partial_{x'}\Theta_{R}(x') \nonumber\\& &
\end{eqnarray}
\begin{eqnarray}
&&H^{n\neq 0}_{s}(\Theta_{s},\Phi_{s};\hat{\alpha}_{s},\hat{Q}_{s},l(V_{G}))= \hbar v_{s,R}(l(V_{G}))[\int_{-L/(2e^{l(V_{G})})}^{L/(2e^{l(V_{G})}) }\,dx[\frac{K_{s}(l(V_{G}))}{2}(\partial_{x}\Phi_{s,R}(x))^{2}\nonumber\\&&+\frac{1}{2K_{s}(l(V_{G})}(\partial_{x}\Theta_{s,R}(x))^{2}] +g_{s}(l(V_{G})\cos(\sqrt{8\pi }\Theta_{s,R}(x)+\hat{\alpha}_{s}+\frac{2\pi}{L}\hat{Q_{s}}x e^{l(V_{G})})]]
\end{eqnarray}
Since $K_{s}(l(V_{G}))\geq1$,  the Sine-Gordon scaling shows that $g_{s}(l(V_{G}))$ is an irrelevant coupling constant which  decreases with the increase of  $l(V_{G})$.
The renormalized  zero mode  Hamiltonian will depend on the  renormalized coupling constants given by the R.G. eqs. $(9-13)$:
\begin{equation}
:H^{(n=0)}(l(V_{G})):= :H^{(n=0)}_{0}(l(V_{G})):+H^{(n=0)}_{int.}(l(V_{G})):
\label{zerozero}
\end{equation}
The first term $ :H^{(n=0)}_{0}(l(V_{G})):$ represents   the non-interacting part:
\begin{equation}
:H^{(n=0)}_{0}(l(V_{G})):=\frac{h v_{F}}{2L}[\hat{N}^{2}_{R,\sigma=\uparrow}+\hat{N}^{2}_{L,\sigma=\uparrow}+\hat{N}^{2}_{R,\sigma=\downarrow}+\hat{N}^{2}_{L,\sigma=\downarrow}]
\label{wire}
\end{equation}
 The second term represents the interactions    :$H^{(n=0)}_{int.}(l(V_{G})):$, given as a function of the charge operator  $ Q_{c}= [(N_{R,\sigma=\uparrow}+N _{L,\sigma=\uparrow})+(N_{R,\sigma=\downarrow}+N _{L,\sigma=\downarrow})]=\hat{Q}_{c}+ <F|Q_{c}|F>$  and the magnetization operator     $Q_{s}=[(N_{R,\sigma=\uparrow}+N _{L,\sigma=\uparrow})-(N_{R,\sigma=\downarrow}+N _{L,\sigma=\downarrow})]= \hat{Q}_{s}+<F|Q_{s}|F>$. 
\begin{equation}
:H^{(n=0)}_{int.}(l(V_{G})):=\eta_{c}(l(V_{G}))\hat{Q}^2_{c}-\eta_{s}(l(V_{G}))\hat{Q}^2_{s}
\label{int}
\end{equation}
where    $\eta_{c}(l(V_{G}))\equiv \frac{h v_{F}}{2L}[(\frac{1-K^{2}_{R}(l(V_{G})}{K^{2}_{R}(l(V_{G}))})+\gamma (\frac{c}{v_{F}})F(\frac{L}{d e^{l(V_{G})}},\frac{\xi}{d e^{l(V_{G})}})]$   are  the renormalized backward  charge and magnetic  interactions 
   $\eta_{s}(l(V_{G}))= \frac{h v_{F}}{2L}(\frac{1-K^{2}_{s,R}(l(V_{G}))}{K^{2}_{s,R}(l(V_{G}))})$. 
Both terms are a function of the screened Coulomb interaction $F(\frac{L}{d e^{l(V_{G})}},\frac{\xi}{d e^{l(V_{G})}})$  given by:  
\begin{equation}
F(\frac{L}{d e^{l(V_{G})}},\frac{\xi}{d e^{l(V_{G})}})=log[\frac{\sqrt{[1+(\frac{d e^{l(V_{G})}}{L})^2]}+1}{\sqrt{[1+(\frac{d e^{l(V_{G})}}{L})^2]}-1}]-log[\frac{\sqrt{[1+(\frac{\xi e^{l(V_{G})}}{L})^2]}+1}{\sqrt{[1+(\frac{\xi e^{l(V_{G})}}{L})^2]}-1}]
\label{log}
\end{equation}
At finite temperatures the effect of the e-e interactions replaces the non-interacting ground state $|F>$ with a shifted Fermi surface given by the renormalized ground state  $|G>$ .  In Appendix $B$ we find that the single particles states $\epsilon(n)$ are shifted up in energy by the self energy   $\delta\Sigma (V_{G},l(V_{G}))$.
In Appendix $B$ we have computed  the   self energy $\delta\Sigma (V_{G},l(V_{G}))$ at  low temperatures $T$ which are higher than the spin exchange   energy, $K_{B}T>\eta_{s}(l(V_{G}))<G|\hat{Q}^2_{s}|G>= K_{B}T^{*}$ .  $|G>$ is the renormalized Fermi Surface which replaces the non - interacting Fermi surface $|F>$ and $T^{*}$ is a temperature of the order of $0.05$ Kelvin. For temperatures $T> T^{*}$  the self energy   is given by 
  $\delta\Sigma (V_{G},l(V_{G}))\approx2\eta_{c}(l(V_{G}))<G|\hat{Q}_{c}|G>$   (see Appendix $B$).
The effective zero mode Hamiltonian is replaced by:
\begin{eqnarray}
&&:H^{(n=0)}(l)_{eff}:\approx \frac{h v_{F}}{2L}[\hat{N}^{2}_{R,\sigma=\uparrow}+\hat{N}^{2}_{L,\sigma=\uparrow}+\hat{N}^{2}_{R,\sigma=\downarrow}+\hat{N}^{2}_{L,\sigma=\downarrow}]\nonumber\\&&+\delta\Sigma (V_{G},l(V_{G}))[\hat{N}_{R,\sigma=\uparrow}+ \hat{N}_{R,\sigma=\downarrow}+\hat{N} _{L,\sigma=\uparrow}+\hat{N}_{L,\sigma=\downarrow}]
\label{eqnarray}
\end{eqnarray}

\vspace{0.1 in}

\textbf{7. The current for  the interacting region  $\frac{1}{2} \leq K_{eff.}(l(V_{G})\leq1$,  $T>T^{*}$}

\vspace{0.1 in}

For finite values of $l(V_{G})$ the spin density wave coupling constant  $ g_{s}(l(V_{G})$    and the spin density wave velocity $v_{s}(l(V_{G})=\frac{v_{F}}{K_{s}(l(V_{G})} <<\frac{v_{F}}{K_{c}(l(V_{G})}=v(l(V_{G})$  are   both small.
At  temperatures  $T>T^{*}$, 
we  replace the interacting zero mode Hamiltonian with the effective zero mode Hamiltonian controlled by the    self energy   $\delta\Sigma (V_{G},l(V_{G}))$ given in eq.$(20)$. 
In order to  compute the current, we include the reservoir Hamiltonian $H_{Res}$  controlled by  the drain source voltage $V=\frac{\mu^{(0)}_{Left}-\mu^{(0)}_{Right}}{e}$:
\begin{equation}
H_{Res}= \frac{eV}{2}\sum_{\sigma=\uparrow,\downarrow} [(\hat{N}_{L,\sigma}-\hat{N}_{R,\sigma})]
\label{res}
\end{equation}
The partition functions in the presence of the reservoir is given by:  
$Z=T_{r}[e^{-\beta :H^{(n=0)}(l(V_{G}))_{eff}:} e^{-\beta:H_{Res}:}]$ $\equiv  T_{r}[e^{-\beta :H^{(n=0)}_{0}:}e^{-\beta:H^{eff.}_{Res}:}]$.
The self energy  allows us to replace  the   reservoir  Hamiltonian $H_{Res}$ by an effective reservoir $H^{eff.}_{Res}$ :
\begin{equation}
H^{eff.}_{Res}=H_{Res}+\sum_{\sigma=\uparrow,\downarrow} [\delta\Sigma (V_{G},l(V_{G}))(\hat{N}_{L,\sigma}+\hat{N}_{R,\sigma})]
\label{nres}
\end{equation}
The static  conductivity is  computed  using the non-interacting zero mode Hamiltonian   $H^{(n=0)}_{0}$  given in eq.$(17)$ and  effective reservoir   $H^{eff.}_{Res}$  given by eq.$(22)$.  
The  current is obtained from  the derivative of the zero mode coordinate $\hat{\alpha}$ (see chapter  III ),  $\hat{I}=\frac{e}{2\pi}\frac{d \hat{\alpha}}{dt}$.  Using the  Heisenberg  equation of motion   we obtain the current operator.
\begin{equation}
\hat{I}=\frac{e}{2\pi}\frac{d \hat{\alpha}}{dt}=\frac{e}{i\hbar}[\hat{\alpha},H^{(n=0)}_{0}]=\frac{e v_{F}}{L}\sum_{\sigma=\uparrow,\downarrow}[\hat{N}_{R,\sigma}-\hat{N}_{L,\sigma}]
\label{current}
\end{equation}
The thermal expectation function is obtained with the  help of the  partition function $Z$. 
\begin{equation}
I=T_{r}[e^{-\beta H^{(n=0)}_{0}}e^{-\beta H^{eff.}_{Res}}\hat{I}][Z]^{-1}
\label{opera}
\end{equation}
Following Appendix $A$  we obtain:
 \begin{eqnarray}
I&=&\frac{e v_{F}}{L}\sum_{\sigma=\uparrow,\downarrow}\sum_{m=-n_{F}(V_{G})}^{m=n_{F}(V_{G})}
([f_{F.D.}[\frac{\epsilon_{L}(m)+\delta\Sigma (V_{G},l(V_{G}))+ \frac{eV}{2}-\delta\mu_{0}(T)}{K_{B}T}]\nonumber\\&&-f_{F.D.}[\frac{\epsilon_{R}(m)+\delta\Sigma (V_{G},l(V_{G}))- \frac{eV}{2}-\delta\mu_{0}(T)}{K_{B}T}])
\label{eqnarray}
\end{eqnarray}
where  $\epsilon_{L}(m)$ and  $\epsilon_{R}(m)$ are the single particle energies  and $2n_{F}(V_{G})$ is the discrete bandwidth  introduced in Appendix $A$.
We include  a small single particle broadening  which will allow us to
 replace  the discrete sum  $\epsilon_{R,L}(m)$ by a continuum  integration  variable $\epsilon$. Performing the integration  with respect the energy  variable  $\epsilon$ and expanding  with respect the voltage    $V$ gives   the conductance   $G=\frac{I}{V}$:
\begin{eqnarray}
G&\approx &\frac{2e}{hV}\int_{-\epsilon_{F}(V_{G})}^{\epsilon_{F}(V_{G})}d\epsilon( f_{F.D.}[\frac{\epsilon +\delta\Sigma (V_{G},l(V_{G}))+ \frac{eV}{2} -\delta\mu_{0}(T)}{K_{B}T}]-f_{F.D.}[\frac{\epsilon +\delta\Sigma (V_{G},l(V_{G}))- \frac{eV}{2} -\delta\mu_{0}(T)}{K_{B}T}])\nonumber\\&&=\frac{2e^2}{h}(f_{F.D.}[\frac{-\epsilon_{F}(V_{G}) +\delta\Sigma (V_{G},l(V_{G})) -\delta\mu_{0}(T)}{K_{B}T}]-f_{F.D.}[\frac{\epsilon_{F}(V_{G}) +\delta\Sigma (V_{G},l(V_{G})) -\delta\mu_{0}(T)}{K_{B}T}])
\label{eqnarray}
\end{eqnarray}
We observe that the self energy  determines the conductance through an effective chemical potential. The bottom of the bandwidth  $ -\epsilon_{F}(V_{G}) $  is replaced by  $-\epsilon_{F}(V_{G})+\delta\Sigma (V_{G},l(V_{G}))$. This allows to introduce the renormalized effective chemical potential $\mu_{R}(V_{G},l(V_{G}))=\epsilon_{F}(V_{G})-\delta\Sigma (V_{G},l(V_{G}))$.

\vspace{0.1 in}

\textbf{8. The strongly interacting region    $K_{eff.}(l(V_{G}))\approx \frac{1}{2}$  ,  $T>T^{*}$ - The  effective  $U=\infty$ Hubbard model}   
 
\vspace{0.1 in}

When  $K_{eff.}(l(V_{G}))\approx \frac{1}{2}$  and $T>T^{*}$,   one obtains an incoherent Luttinger liquid  which can be mapped to the  Hubbard model  $U\rightarrow\infty$.   
(When  $U\rightarrow\infty$ the interaction  Luttinger parameter is given by $K\rightarrow\frac{1}{2}$ and the spin excitations which are  of the order $\frac{1}{U}$  can be ignored.)   
This limit $U\rightarrow\infty$  has been considered in the past  \cite{slaveD}. 
In this limit the following constraints must  be obeyed: $\psi^{+}_{\sigma=\uparrow}(x)\psi_{\sigma=\uparrow}(x)+\psi^{+}_{\sigma=\downarrow}(x)\psi_{\sigma=\downarrow}(x)=0,1$.  Using the  constraints, we have  found  the following representation \cite{slaveD} for the electron operators: $\psi_{\sigma}(x)=b_{\sigma}(x)\Psi(x)$, $\psi^{+}_{\sigma}(x)=\Psi^{+}(x)b^{+}_{\sigma}(x)$ where  $\Psi(x)$ is the electron  charge operator and $b_{\sigma}(x)$ are the hard core boson for the spin excitations.   They  obey  the constraints:
 $ b^{+}_{\sigma=\uparrow}(x)b_{\sigma=\uparrow}(x)+ b^{+}_{\sigma=\downarrow}(x)b_{\sigma=\downarrow}(x)=\Psi^{+}(x)\Psi(x)$.  
 In one dimension, this model has been represented    in terms of the Bosonic  electron  operators \cite{slaveD} $\Theta_{e}$  and $\Phi_{e}$ and Spinon  operators  $\Theta_{s}$  and $\Phi_{s}$. The constraint is imposed on the electron density:      $\rho_{e}(x)\equiv\rho_{\sigma=\uparrow}(x)+ \rho_{\sigma=\downarrow}(x)=\frac{1}{\sqrt{\pi}}[\partial_{x}\Theta_{\sigma=\uparrow}(x)+\partial_{x}\Theta_{\sigma=\downarrow}(x)\equiv\frac{1}{\sqrt{\pi}}\partial_{x}\Theta_{e}(x)$. The canonical conjugate momentum is given by:   $\partial_{x}\Phi_{e}(x)\equiv\frac{1}{2}[\partial_{x}\Phi_{\sigma=\uparrow}(x)+ \partial_{x}\Phi_{\sigma=\downarrow}(x)]$.
For non - interacting electrons  we have the commutation rule $[\Theta_{e}(x), \partial_{x}\Phi_{e}(y)]=i\hbar\delta(x-y)$. Due to the exclusion  of  double occupancy,   the  electronic density is reduced  by a factor of two  (in comparison with non - interacting electrons) and the commutator is modified to:  
$[\Theta_{e}(x), \partial_{x}\Phi_{e}(y)]_{Constraint}\approx \frac{i}{2}\hbar\delta(x-y)$.

The Hamiltonian  for the $U\rightarrow\infty$ case   (away from half  filling ) is given  in terms of the fields $\Theta_{e}(x)$, $\Phi_{e}(x)$:
$H_{e}=\int\,dx v\hbar[(\partial_{x}\Phi_{e}(x))^2+ \frac{1}{4}(\partial_{x}\Theta_{e}(x))^2]$.

At finite temperatures, the spinon  Hamiltonian  $H_{s}= \int\,dx \frac{\hbar v}{2}[(\partial_{x}\Phi_{s}(x))^2+ (\partial_{x}\Theta_{s}(x))^2]\approx 0$   is negligible ( $\Theta_{s}(x)\equiv\frac{1}{\sqrt{2}}[ \Theta_{\sigma=\uparrow}(x)-\Theta_{\sigma=\downarrow}(x)]$  and $\Phi_{s}(x)=\frac{1}{\sqrt{2}}[ \Phi_{\sigma=\uparrow}(x)  -\Phi_{\sigma=\downarrow}(x)]$).  If we inject an electron with a given spin at one  lead,   we will detect on the other lead a charge with an arbitrary spin.
The effect of   voltage  difference  $V$ between the leads,  is included into the calculation through     the reservoir Hamiltonian   $\frac{eV}{2\sqrt{ \pi}}\int\,dx[\partial_{x}\Phi_{\sigma=\uparrow}(x)+\partial_{x}\Phi_{\sigma=\downarrow}(x)]\equiv\frac{eV}{\sqrt{ \pi}}\int\,dx\partial_{x}\Phi_{e}(x)$.
Using the Heisenberg equations of motion with the modified commutator we obtain  the  electronic current  operator \cite{constraint}    $J_{e}=\frac{e v}{2\sqrt{\pi}}[\partial_{x}\Phi_{\sigma=\uparrow}(x)+\partial_{x}\Phi_{\sigma=\downarrow}(x)]\equiv \frac{e v}{\sqrt{\pi}} \partial_{x}\Phi_{e}(x)$.
  The extra  factor of $\frac{1}{2}$  which appears in  the current operator  $J_{e}$ is due to the  commutator $[,]_{Constraint}$ \cite{constraint}. Therefore, the  conductance  is reduced     to    $G\approx\frac{e^2}{h}$.

\vspace{0.1 in}

\textbf{9. The   conductance  in the Wigner crystal limit  $K_{eff.}(l(V_{G}))< \frac{1}{2}$} 

\vspace{0.1 in}

For  large    screening lengths $\xi$, the effective Luttinger charge stiffness 
$K_{eff.}(l(V_{G}))$  decreases below    $K_{eff.}(l(V_{G}))<\frac{1}{2}$ at low temperatures.  Under these conditions, the  R.G. analysis reveals that   the  alternating umklapp coupling constant $g(-1)^{n}$   generates a gap at  $T=0$ . 
To investigate this region, we introduce  two  sub - lattices  for the even and odd   sites.  We replace the Bosonic fields by the even and odd combinations  : $\Theta_{-}(y=2nd)=\frac{\Theta(x=2nd)-\Theta(x=(2n+1)d)}{\sqrt{2}}$  and  $\Theta_{+}(y=2nd)=\frac{\Theta(x=2nd)+\Theta(x=(2n+1)d)}{\sqrt{2}}$.  We   integrate out the  antisymmetric field $\Theta_{-}(y=2nd)=\frac{\Theta(x=2nd)-\Theta(x=(2n+1)d)}{\sqrt{2}}$) and obtain  an effective Hamiltonian   for the  symmetric field  $\Theta_{+}(y=2nd)=\frac{\Theta(x=2nd)+\Theta(x=(2n+1)d)}{\sqrt{2}}$.  The effective Hamiltonian has a set of   new coupling constants  $|g^{(2)}_{new}|\approx  \frac{g^2}{2!} <(\sin[\sqrt{4\pi }\Theta_{-}(y)])^2>$. 
\begin{eqnarray}
H^{n\neq 0}_{c}(\Theta_{+},\Phi_{+})&&\approx \hbar v(l(V_{G}))[\int_{-L/2}^{L/2}\,dy[\frac{K_{eff.}(l(V_{G}))}{2}(\partial_{y}\Phi_{+}(y))^{2}+\frac{1}{2K_{eff.}(l(V_{G}))}(\partial_{y}\Theta_{+}(y))^{2} \nonumber\\&&+g^{(2)}_{new,R}(l(V_{G}))\cos(\sqrt{16\pi }\Theta_{+}(y))]] 
\end{eqnarray}
where   the new coupling constant  $g^{(2)}_{new,R}(l(V_{G}))$ obeys the R.G. equation. 
\begin{equation}
\frac{d\hat{g}^{(2)}_{new,R}(l)}{dl}= 
   2 \hat{g}^{(2)}_{new,R}(l)[1 -  2K_{eff.}(l)]
\label{newscaling}
\end{equation}
This equation shows that  $\hat{g}^{(2)}_{new,R}(l)$ is a relevant coupling constant  for  $K_{eff.}(l(V_{G})<\frac{1}{2}$.  As a result, a charge  gap  $\Delta\approx \Lambda(\hat{g}_{new}^{(2)})^{\frac{1}{2( 2K_{eff.}(l(V_{G}))-1)}}$  will   open.
When  $K_{eff.}(l(V_{G})<\frac{1}{2}$,  we obtain from eq.$(28)$ that at 
  $T=0$  the expectation value of the phase   $\sqrt{16 \pi}<\Theta_{+}(x)>=\pi$ will give rise to a Wigner crystal order $\rho(x)\approx constant + cos[4k_{F}x+\frac{\pi}{\sqrt{2}}]e^{-\frac{\pi}{2}<(\Theta(x)-<\Theta(x)>)^2>}$.  (Expanding the cosine term in eq.$(27)$  around the  ground state $<\Theta_{+}(x)>\neq0$ shows that the charge density wave  has a  gap $\Delta$. This gap   suppresses the fluctuations  $e^{-\frac{\pi}{2}<(\Theta(x)-<\Theta(x)>)^2>}\neq0$ and  stabilizes  the   Wigner Crystal order at $T=0$.)
 In order to evaluate the effect of the charge  gap   $\Delta$  on the electronic spectrum  we    map  \cite{Luther,Mori} the Bosonic charge Hamiltonian  to a spinless Fermion for $K_{eff.}(l(V_{G})\approx\frac{1}{2}$. We introduce a two component spinless Fermion:  
 $\chi^{+}(x)=[ \chi_{1}(x), \chi_{2}(x)]^{+}\equiv\sqrt{\frac{\Lambda}{2\pi}}[e^{i\sqrt{4\pi}\Theta_{+}(x)}, e^{-i\sqrt{4\pi}\Theta_{+}(x)}]^{+}$. 
 As a result we find for $K_{eff.}(l(V_{G})\approx\frac{1}{2}$ that the Hamiltonian in  eq. $(27)$ is mapped to a spinless Fermion model:
\begin{eqnarray}  
&&H_{c,F}= \int\,dx[\hbar v(l(V_{G}))\chi^{+}(x)(-i\partial_{x}\sigma_{3})\chi(x)+\frac{\hat{g}^{(2)}_{new,R}(l(V_{G}))}{2}(\chi^{+}(x)\sigma_{1}\chi(x))^2]\nonumber\\&& \approx
\int\,dx[\hbar v(l(V_{G}))\chi^{+}(x)(-i\partial_{x}\sigma_{3})\chi(x)+\hat{g}^{(2)}_{new,R}(l(V_{G}))<\chi^{+}(x)\sigma_{1}\chi(x)>\chi^{+}(x)\sigma_{1}\chi(x)]\nonumber\\&&
\end{eqnarray}
where $\sigma_{3}$ and $\sigma_{1}$ are the Pauli matrices.  As a result, we  have a gap $2\hat{\Delta}$  between the lower band and the  upper band given by the   self consistent solution:
 $\Delta\approx \hat{\Delta}=\hat{g}^{(2)}_{new,R}(l(V_{G}))<\chi^{+}(x)\sigma_{1}\chi(x)>$. 
The energy difference $\Delta$  between the Fermi energy and the top of the lower  electronic band will affect the conductance through the Fermi - Dirac function.   For this case, the self energy is replaced by the gap  $\Delta$ and  the conductance    is approximated by   $G\approx\frac{2e^2}{h}[1-f_{F.D.}(\frac{\Delta}{K_{B}T})]$,   for  $K_{B}T\geq \Delta$  conductances  is  given by  $G\approx  \frac{e^2}{h}$.

\vspace{0.1 in}

\textbf{10. Numerical  results}

\vspace{0.1 in}

We have  used  the experimental relation between the Fermi momentum and the gate voltage $V_{G}$ given by  $K_{F}(V_{G})=\frac{\pi}{2}n_{e}(V_{G})=\frac{C_{a}}{e}(V_{G} -V^{th})$ , $V^{th}\approx-5.52 volt$, $\frac{C_{a}}{e}=1.2\cdot 10^{8} (Volt\cdot meter)^{-1}$  \cite{Picciotto}   to compute   the conductance in figures $2$ and $3$. 
In figure $2$ we have  considered a typical screening ratio  $\frac{ \xi}{d}=10$ and plotted  the conductance for a varying range of temperatures $1-3$ Kelvin.
Figure $3$ shows the conductance at a fixed temperature $T=1$  Kelvin for a different  screening lengths.
We  observe that  for $\frac{ \xi}{d}=1$, the Coulomb  interaction is completely screened and the $0.7$  feature is absent.
In figure $4$,   we plot the dependence of the self energy  $ \delta\Sigma (V_{G},l(V_{G}))$ on the gate voltage $V_{G}$. 
We observe that at low densities 
 the free energy    has an extremum  at a finite density.  The renormalized chemical potential  $\mu_{R}(V_{G},l(V_{G}))=\epsilon_{F}(V_{G})-\delta\Sigma (V_{G},l(V_{G}))$ vanishes at the  voltage  $V^{*}_{G}>V^{th}$.  For  $V^{th}<V_{G}< V^{*}_{G}$,  the renormalized chemical potential is negative, indicating the formation of a   charge density wave gap at $T=0$ for $K_{eff.}(l(V_{G})<\frac{1}{2}$.
The derivative of the  conductance and  chemical potential are  related to the inverse compressibility: 
 $\frac{d G(V_{G})}{d V_{G}}\propto\frac{d \mu_{R}(V_{G},l(V_{G}))}{d V_{G}}\propto \frac{1}{n^2_{e}(V_{G})\kappa(V_{G})}$. 
   The $0.7$   anomaly is  translated into a  minimum  around  $V_{G}=-5.49$ volt for the conductance derivative  $ \frac{d G(V_{G})}{d V_{G}}$ and the compressibility $\kappa (V_{G})$ which is proportional to the inverse square of the effective interaction  parameter $K_{eff.}(l(V_{G}))$  shown in figure $(1)$.   Therefore, we have the  confirmation for the formation of a charge density wave gap for $K_{eff.}(l(V_{G})),\frac{1}{2}$  at zero temperature. 
In figure $5$, we plot the function $\frac{d \mu_{R}(V_{G},l(V_{G}))}{d V_{G}}$.  This function has a minimum at the voltage  $V_{G}=-5.49$ volts, which  corresponds to the $0.7\frac{2e^2}{h}$ structure observed for the  conductance  graph.

\vspace{0.3 in}

\textbf{11. The  effect of the  Zeeman  magnetic  field}

\vspace{0.1 in}

The  $Zeeman$ $magnetic$ field \cite{Pepperoni} introduce  a bias term $(2(k^{\uparrow}_{F}(V_{G})-(k^{\downarrow}_{F}(V_{G}))x $  into the last term in eq.(5).
Expressing the bias in terms of the magnetic field  $B_{||}$ we find, $2(k^{\uparrow}_{F}(V_{G})-(k^{\downarrow}_{F}(V_{G})=2k_{F}(V_{G})(\sqrt{1+\frac{\Delta_{z}}{2\mu_{F}}}-\sqrt{1-\frac{\Delta_{z}}{2\mu_{F}}})\approx 4 k_{F}(V_{G}) (\frac{\Delta_{z}}{2\mu_{F}})$ where
$\mu_{F}=\frac{\hbar^{2}k^2_{F}(V_{G})}{2m^{*}}$ is the Fermi energy  and $\Delta_{z}=g_{||}\mu_{B}B_{||}$ is the Zeeman energy. 
As a result  the spin part  Sine-Gordon term vanishes since $K_{s}(l)>1$. As a result      the spin wave velocity $v_{s}=\frac{v_{F}}{ K_{s}}$ is further   reduced. For large magnetic fields $\frac{\Delta_{z}}{2\mu_{F}}>1$, the wire will be polarized and we will have only one propagating  channel with the conductance  $G\approx 0.5\times(2e^{2}/h)$.

The effect on the charge density wave Hamiltonian will  be to replace   $4 k_{F}(V_{G})$   in eq. $(4)$ by:
$2[k^{\uparrow}_{F}(V_{G})+k^{\downarrow}_{F}(V_{G})]\equiv4k_{F}(V_{G})[\sqrt{1+\frac{\Delta_{z}}{2\mu_{F}}}+\sqrt{1-\frac{\Delta_{z}}{2\mu_{F}}}]\frac{1}{2}$.    
 For $\frac{\Delta_{z}}{2\mu_{F}}<1$ we show that the perfect conductance in the absence of the Zeeman magnetic field computed at the  gate voltage  $V^{0}_{G}$ is shifted  to a larger  gate voltage   $V^{0-Zeeman}_{G}$ in the presence of the Zeeman field:
 $2[k^{\uparrow}_{F}(V^{0-Zeeman}_{G})+k^{\downarrow}_{F}(V^{0-Zeeman}_{G})]\approx 4 k_{F}(V^{0-Zeeman}_{G})[1-\frac{1}{8} (\frac{\Delta_{z}}{2\mu_{F}})^2]= 4 k_{F}(V^{0}_{G})=\frac{\pi}{d}$.
This formula shows the shift in the perfect conductance from  $4k_{F}(V^{0}_{G})=\frac{\pi}{d}$  to a larger gate   voltage $V^{(0-Zeeman)}_{G}>V^{0}_{G}$  given by 
 $k_{F}(V^{(0-Zeeman)}_{G})\approx \frac{k_{F}(V^{0}_{G})}{1-\frac{1}{8}(\frac{\Delta_{z}}{2\mu_{F}})^{2}}$. This result is in   agreement with the experimental observations \cite{Pepperoni}.

The conductance at finite temperatures $T>T^{*}$   will be given  by replacing the self energy in eq. $(26)$  with a new self energy  computed in the presence of the magnetic field, $\epsilon_{F,\sigma=\uparrow}(V_{G})=\epsilon_{F}(V_{G})+\frac{\Delta_{z}}{2}$ and   $\epsilon_{F,\sigma=\downarrow}(V_{G})=\epsilon_{F}(V_{G})-\frac{\Delta_{z}}{2}$. 
The results for the conductance are shown in figure $6$.  We show three graphs:  the first graph (thin line) represents the conductance in the absence of the magnetic field and the other two graphs represent the conductance for the magnetic fields $B=3$ $Tesla$ and $B=10$ $Tesla$. We observe  the shift  of the conductance to higher voltages  with the increase of the  magnetic field .

\vspace{0.1 in}

\textbf{12. Conclusion}

\vspace{0.1 in}

We have presented a model  which  explains  the conductance anomaly at finite temperatures as a function of the gate voltage.  Due to the Coulomb long range interactions a weakly  interacting electronic system can flow to the  strong coupling limit $K_{eff.}(l(V_{G}))<\frac{1}{2}$. 
When the screening length is not too large,  the Luttinger stiffness is restricted to
 $\frac{1}{2}\leq K_{eff.}(l(V_{G}))< 1$ . As a result, the conductance  of an infinite wire is perfect at   zero temperature. At temperatures larger than the magnetic exchange  energy  $T>T^{*}$, we have an incoherent Luttinger model.  For  $K_{eff.}(l(V_{G}))\approx \frac{1}{2}$ the interacting wire  is  equivalent  to the Hubbard $U\rightarrow\infty$ model with the anomalous conductance   $G\approx\frac{e^2}{h}$ .
 
For large screening lengths the interacting charge stiffness   decreases to  $K_{eff.}(l(V_{G}))< \frac{1}{2}$ . As a result we find that at zero temperature we have a  Wigner crystal with a  charge gap  $\Delta$.   At finite temperatures  the formation of charge density wave gap   gives rise to the   anomalous   conductance   $G\approx\frac{2e^2}{h}[1-f_{F.D.}(\frac{\Delta}{K_{B}T})]$. 
Following \cite{Pepperoni}  we have investigated  the  effect of the  magnetic field.  We have shown that the magnetic field  shifts  the region of  the perfect conductance   to higher voltages.

Some of the concepts  used in our work   are common to other theories  \cite{Rech,Michael}.  The long range interactions have been introduced  by  \cite{Matveev, Meyer}; in the present paper we show that, by varying the  gate voltage and screening length we  obtain either a strongly interacting metal    or a Wigner crystal.     
Other theories use   a weak scattering potential  \cite{Reilly}  or  charge localization  \cite{Meir,Meirke} and are consistent with our picture.  
In our view the  origin of the    scattering potential  (microscopic or phenomenological) is not crucial!  
The  crucial effect is that any weak scattering is strongly enhanced  by   the long range interactions!
Our findings show that, due to the long range interaction, any negligible scattering  potential is enhanced and eventually can drive the system to an insulating regime.
It is the interplay between the screening length, gate  voltage and temperature which gives rise to the conductance anomaly.
One of the popular theories is based on the   the Kondo model \cite{Cronennwett,Meir}.   The $Kondo$ picture  dictates that the anomaly should be observed above the Kondo temperature. When the temperature is  lowered  below the  Kondo temperature, the conductance is restored to the universal value.
This picture is  consistent with our theory in the following way:
If  the strong coupling  regime  $K_{eff.}(l(V_{G}))\approx \frac{1}{2}$  is reached in the metallic phase, we can use the Hubbard $U\rightarrow\infty$ limit,  which is the basis  for deriving  the Kondo model.  The Kondo  physics emerges for finite exchange coupling $J\propto \frac{1}{U}$ . In our case, the anomalous conductance is observed at finite temperatures   $T>T^{*}$ which is comparable to  $\frac{1}{U}$  where the Kondo  picture emerges.

\textbf{Acknowledgements}:
D.Shmeltzer wants to   thank  Dr. Jing Qiao Zhang for his invaluable help  and guidance with the  computational part  and graphical presentation  of this work.
The authors  acknowledge the financial support from the CUNY  Collaborative  Grant   award  for the year 2007-2008.

\clearpage

\begin{figure}
\begin{center}
\includegraphics[width=6.5 in ]{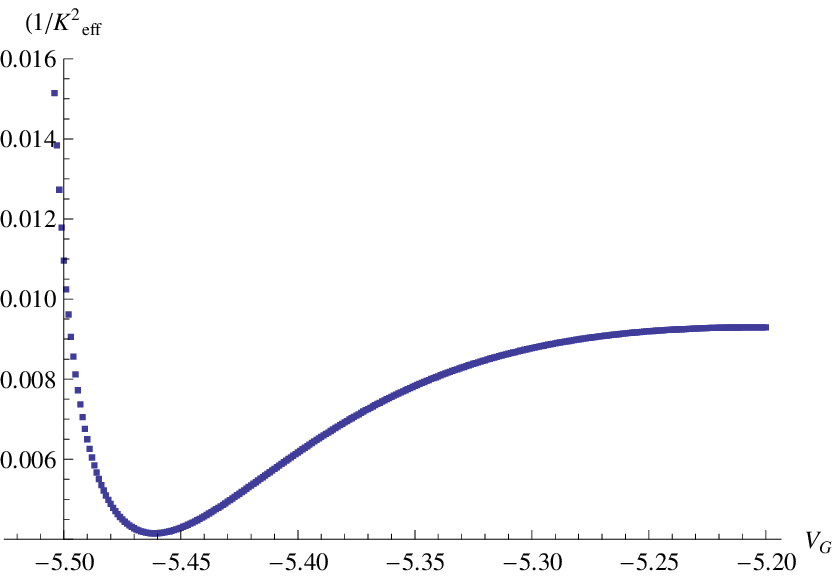}
\end{center}
\caption{ The  effective parameter  $\frac{1}{K^2_{eff.}(l(V_{G})}$  that is proportional  to the inverse compressibility is plotted   as a function  of the   gate voltage $l=l(V_{G})$ for, $L= 10^{-6}$ meter and the  screening ratio $\frac{ \xi}{d}=10$}
\end{figure}

\begin{figure}
\begin{center}
\includegraphics[width=6.0 in ]{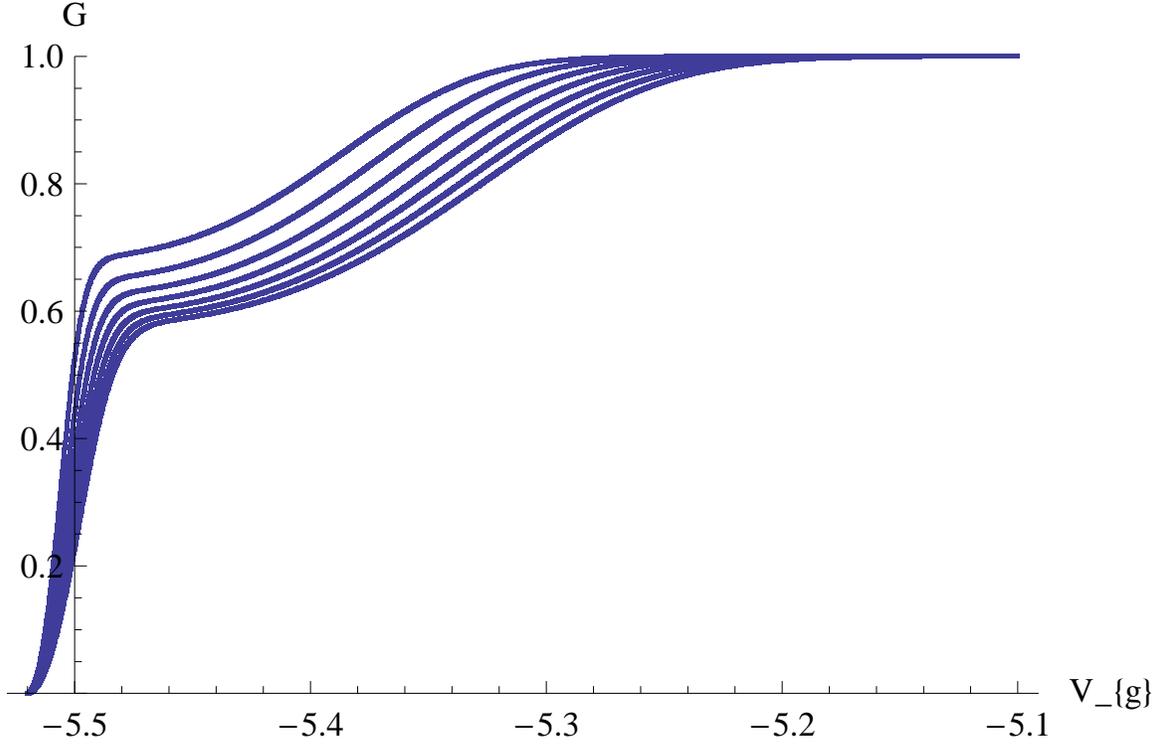}
\end{center}
\caption{The Conductance $G$ in units of $\frac{2e^2}{h}$  as a function of the bias  gate voltage $l=l(V_{G})$ for the  temperatures $T=1.$ $Kelvin$ (upper line),  $T=1.25$ $Kelvin$, $T=1.5$ $Kelvin$,$T=1.75$ $Kelvin$,$T=2.0$ $Kelvin$,$T=2.25 Kelvin$,$T=2.5 Kelvin$ and  $T=3.$ $Kelvin$  (the lowest  line) for umklapp parameter $g(l=0)=0.05$, $K(l=0)\approx0.98$ ,$L=10^{-6}$ $m$  and  screening ratio $\frac{ \xi}{d}=10$}
\end{figure}

\begin{figure}
\begin{center}
\includegraphics[width=6.0 in ]{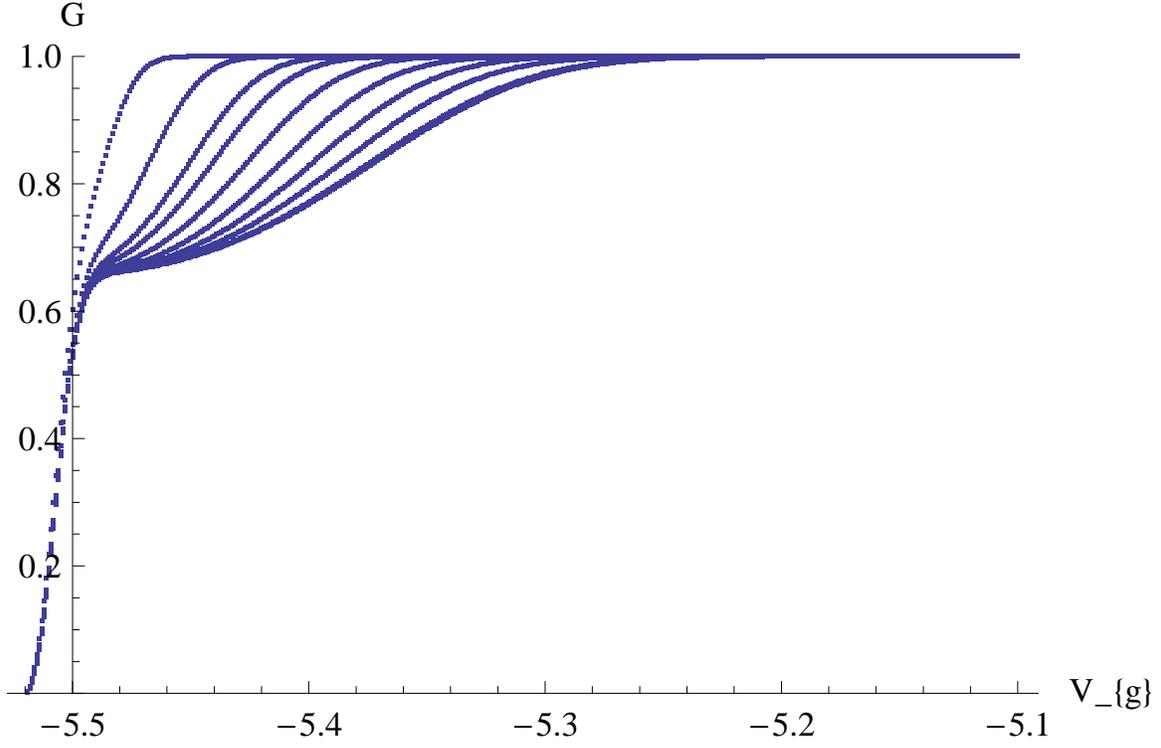}
\end{center}
\caption{The Conductance $G$ in units of $\frac{2e^2}{h}$   for four screening ratios  $\frac{ \xi}{d}=1$(upper line), $\frac{ \xi}{d}=1.1$,  $\frac{ \xi}{d}=1.3$, $\frac{ \xi}{d}=1.5$, $\frac{ \xi}{d}=2.$, $\frac{ \xi}{d}=3.$, $\frac{ \xi}{d}=5.$, $\frac{ \xi}{d}=10$,$\frac{ \xi}{d}=50.$ and $\frac{ \xi}{d}=100$ at  temperature $T=1.$ $Kelvin$   length $L=10^{-6}$ $m$   for the interactions parameters  $\hat{g}_{R}(l=0)=0.05$, $K(l=0)\approx 0.98$ }
\end{figure}
 
 \begin{figure}
\begin{center}
\includegraphics[width=7.0 in ]{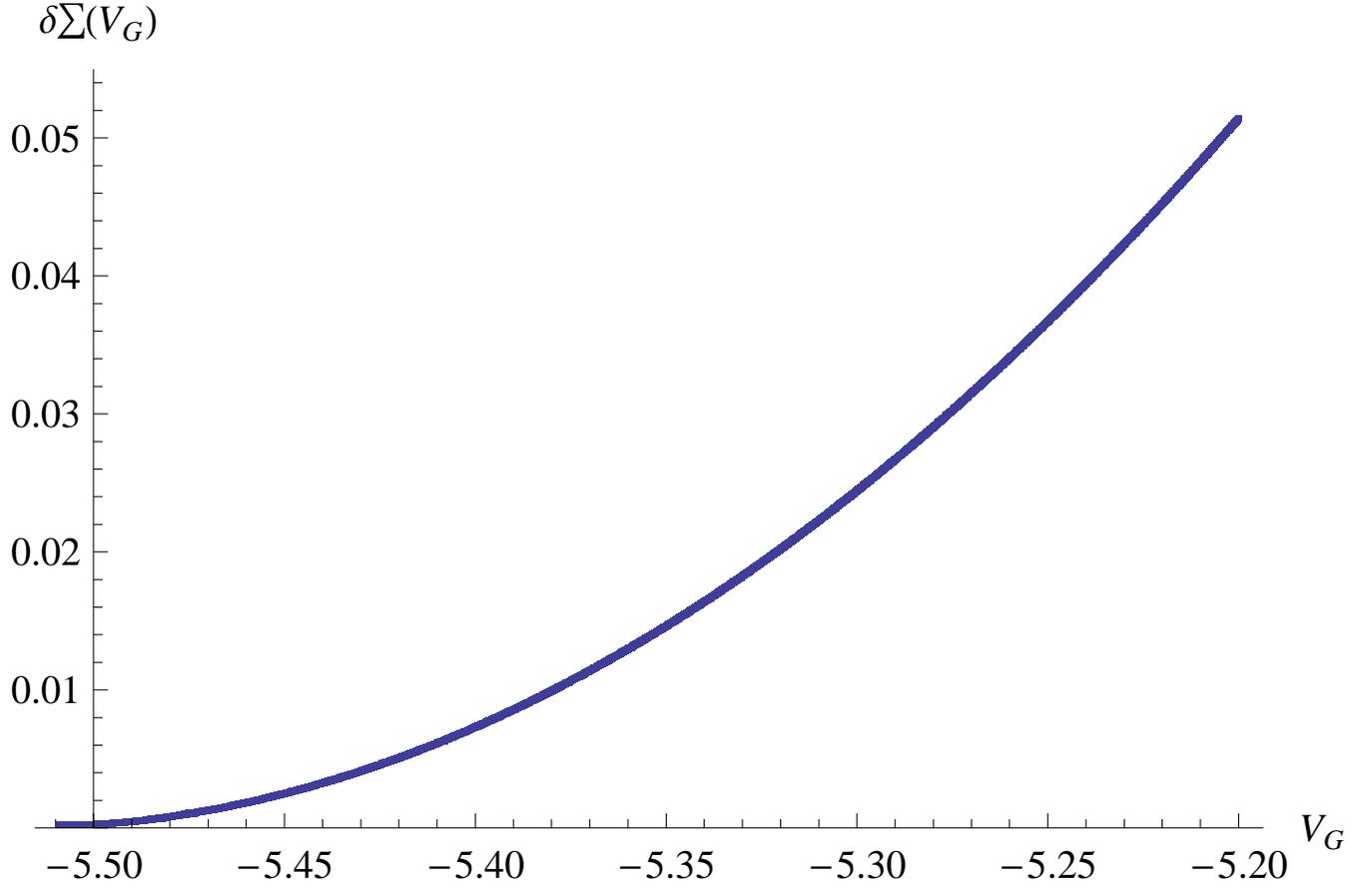}
\end{center}
\caption{The shift in the  chemical potential  $ \delta\Sigma (V_{G},l(V_{G}))$  for  screening ratio  $\frac{ \xi}{d}=10$ at  temperature $T=1. Kelvin$   length $L= 10^{-6} m$   for the interactions parameters  $\hat{g}_{R}(l=0)=0.05$, $K(l=0)\approx 0.98$ }
\end{figure}

\begin{figure}
\begin{center}
\includegraphics[width=7.0 in ]{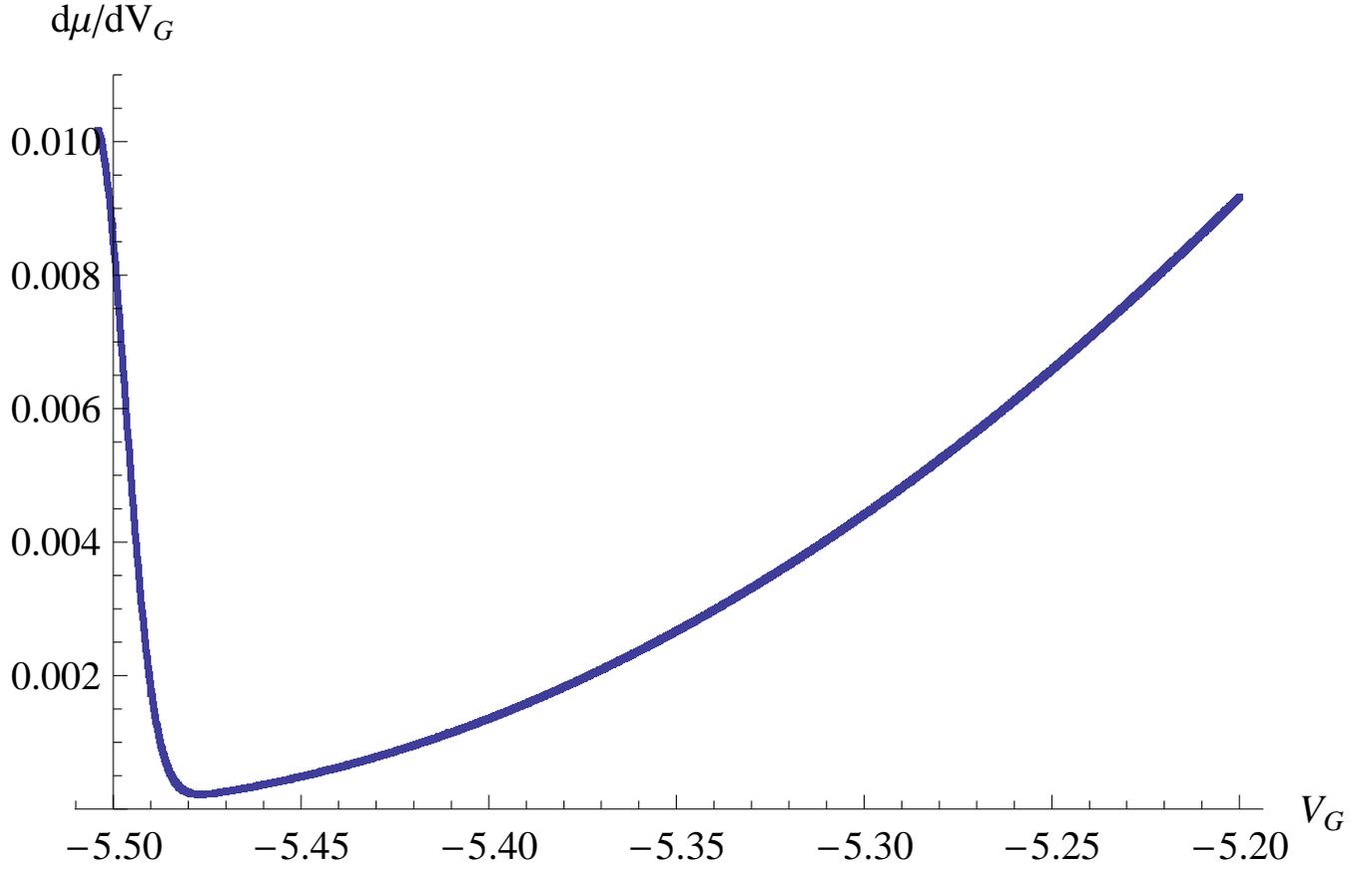}
\end{center}
\caption{The derivative of the chemical potential  $\frac{d \mu_{R} (V_{G},l(V_{G}))}{dV_{G}}$  for  screening ratio  $\frac{ \xi}{d}=10$ at  temperature $T=1. Kelvin$   length $L= 10^{-6} m $   for the interactions parameters  $\hat{g}_{R}(l=0)=0.05$, $K(l=0)\approx 0.98$ }
\end{figure}

\begin{figure}
\begin{center}
\includegraphics[width=6.5 in ]{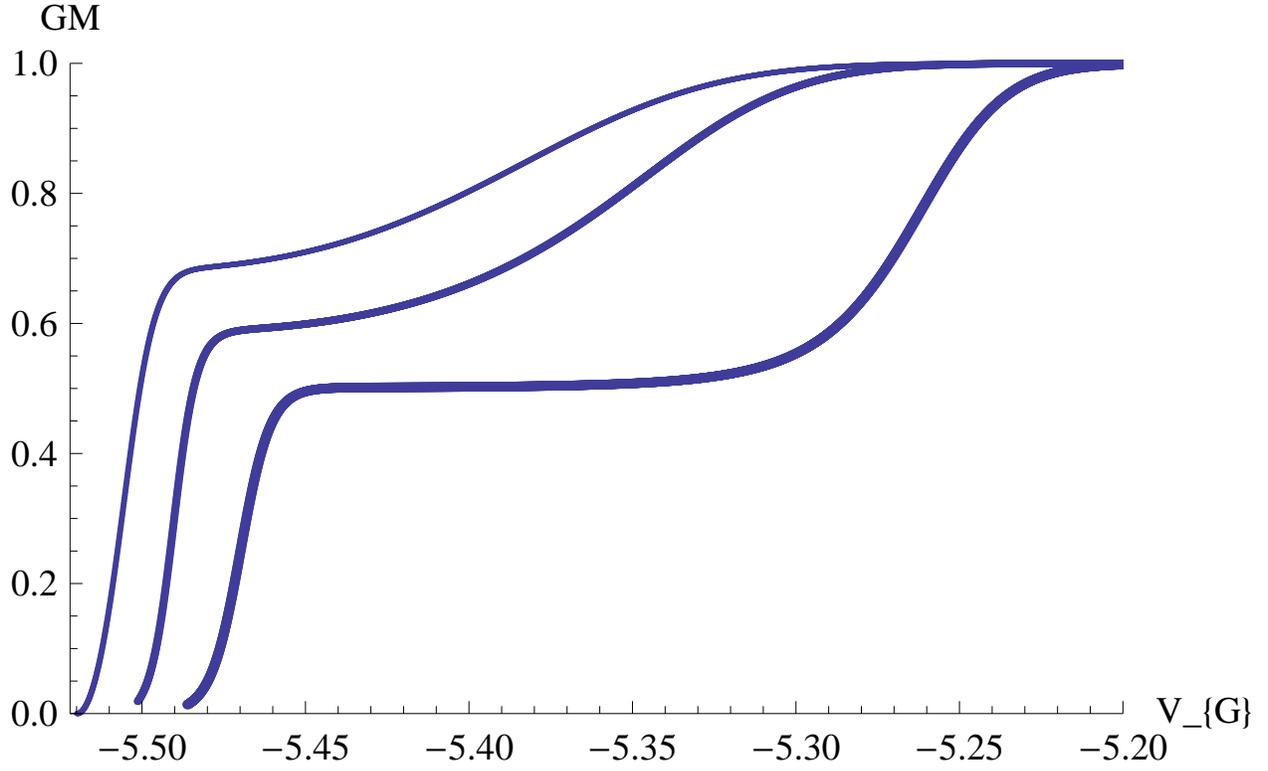}
\end{center}
\caption{ The  effect of the   magnetic field on the  conductance  in units   ${2e^2}{h}$.    The first graph represents the conductance for zero magnetic field, the second graph represents the conductance for a magnetic field $B=3$ $Tesla$  and the third graph represents the conductance for the magnetic field  $B=10$ $Tesla$.  The other parameters were: screening ratio  $\frac{ \xi}{d}=10$, temperature $T=1$ $Kelvin$,  length  $L= 10^{-6}$  $m$, $d=10^{-8}$ $m$, $\hat{g}_{R}(l=0)=0.05$ and $K(l=0)\approx 0.98$}  
\end{figure}
\clearpage

\textbf{Appendix-A}

 The  non-interacting Fermi surface at $T=0$ is given by the state $|F>$,   which is constructed from the vacuum   $|0>$:
$|F>\equiv \prod_{\sigma=\uparrow,\downarrow}[\prod_{-n_{F}(V_{G})}^{n_{F}(V_{G})}R^{+}{(m,\sigma)}\prod_{n_{F}(V_{G})}^{-n_{F}(V_{G})}L^{+}{(m,\sigma})]|0>$. 
 We introduce the notation  $\hat{N}_{R,\sigma}$ $\hat{N}_{L,\sigma}$ for the normal order at zero temperature:
\begin{equation}
\hat{N}_{R,\sigma}=\sum_{-n_{F}(V_{G})}^{n_{F}(V_{G})}R^{+}{(m,\sigma)}R{(n,\sigma)}-\sum_{-n_{F}(V_{G})}^{n_{F}(V_{G})}<F|R^{+}{(m,\sigma)}R{(m,\sigma)}|F>\equiv N_{R,\sigma}-<F|N_{R,\sigma}|F>
\label{right}
\end{equation}
\begin{equation}
\hat{N}_{L,\sigma}=\sum_{n_{F}(V_{G})}^{-n_{F}(V_{G})}L^{+}{(m,\sigma)}L{(m,\sigma)}-\sum_{n_{F}(V_{G})}^{-n_{F}(V_{G})}<F|L^{+}{(m,\sigma)}L{(m,\sigma)}|F>=N_{L,\sigma}-<F|N_{L,\sigma}|F>
\label{left}
\end{equation} 
The presence of  a reservoir with two chemical potentials  $\mu_{R}$ and $\mu_{L}$  is described by the reservoir Hamiltonian:
\begin{equation} 
H_{Res}= \mu_{R}\sum_{-n_{F}(V_{G})}^{n_{F}(V_{G})}R^{+}{(m,\sigma)}R{(m,\sigma)} +\mu_{L}\sum_{-n_{F}(V_{G})}^{n_{F}(V_{G})}L^{+}{(m,\sigma)}L{(m,\sigma)}
\label{substraction}
\end{equation}
At finite temperatures,  the Fermi surface is shifted  by $\delta\mu_{0}(T)$  and is given (for the one dimensional case) by:
$\delta\mu_{0}(T)=\epsilon_{F}(V_{G})\frac{\pi^2}{12}(\frac{K_{B}T}{\epsilon_{F}(V_{G})})^2$). The temperature and the reservoir modifies the number of fermion in the thermal ground state to $<N_{R,\sigma}(V_{G},\mu_{R},T)>$ and  $<N_{L,\sigma}(V_{G},\mu_{L},T)>$ given by:
\begin{equation}
<N_{R,\sigma}(V_{G},\mu_{R},T)>=\sum_{n_{F}(V_{G})}^{-n_{F}(V_{G})}f_{F.D.}[\frac{\epsilon_{R}(m) -\mu_{R}-\delta\mu_{0}(T)}{K_{B}T}]
\label{equation1}
\end{equation}
\begin{equation}
<N_{L,\sigma}(V_{G},\mu_{L},T)>=\sum_{n_{F}(V_{G})}^{-n_{F}(V_{G})}f_{F.D.}[\frac{\epsilon_{L}(m) -\mu_{L}-\delta\mu_{0}(T)}{K_{B}T}]
\label{equation2}
\end{equation}
The expectation value of the normal order operators will be given by: 

 $<\hat{N}_{L,\sigma}(V_{G},\mu_{L},T)>=<N_{L,\sigma}(V_{G},\mu_{L},T)>-<N_{L,\sigma}(\delta\mu_{0}(T),T)>$;
 
$<\hat{N}_{R,\sigma}(V_{G},\mu_{R},T)>=<N_{R,\sigma}(V_{G},\mu_{R},T)>-<N_{R,\sigma}(\delta\mu_{0}(T),T)>$.

The effect  of the self energy     $\delta\Sigma (V_{G},l(V_{G}),T)$ will be   taken in consideration by substituting  in the previous equations  :     $\mu_{R}\rightarrow   \mu_{R}-\delta\Sigma (V_{G},l(V_{G}),T)$ and 
 $\mu_{L}\rightarrow   \mu_{L}-\delta\Sigma (V_{G},l(V_{G}),T)$ .

\textbf{Appendix-B}

The purpose of this Appendix is to  compute the self energy  $\delta\Sigma (V_{G},l(V_{G}),T)$ for the following model:
\begin{equation}
:H^{(n=0)}_{int.}(l):=\eta_{c}(l(V_{G}))\hat{Q}^2_{c}-\eta_{s}(l(V_{G}))\hat{Q}^2_{s}
\label{int}
\end{equation}
where $Q_{c}= \hat{Q}_{c}+ <F|Q_{c}|F>$ is the charge operator and  $Q_{s}= \hat{Q}_{s}+ <F|Q_{s}|F>$ is  the magnetization operator.
We observe that  the zero mode  component of the Hamiltonian  commutes: $[H^{(n=0)}_{0}(l),H^{(n=0)}_{int.}(l)]=0$.   Therefore, at finite temperatures, the 
partition function   $Z ^{(n=0)}=T_{r}[e^{-\beta H^{(n=0)}(l)}]$ can be  computed exactly.
Our goal is to compute the charge current $\hat{I}=e\frac{d \hat{\alpha}}{dt}$,
which is given  by  the commutator $[\hat{\alpha},H^{(n=0)}(l)]$.
We will limited ourselves to finite temperatures such that the exchange  energy  is smaller than the thermal energy and therefore, can be ignored (for long wires the spin stiffness approaches  $K_{s}\approx 1$ and the the last term in eq.$(19)$ $\eta_{s}(l(V_{G}))$ vanishes).   
We will compute  the self energy  at finite temperature  $\delta\Sigma (V_{G},l(V_{G}))$.
For the non-interacting ground state $|F>$ with the electronic density $n_{e}(V_{G})$ we have at a temperature $T$ the equation: $n_{e}(V_{G})=
\frac{<F|Q_{c}|F>}{L}=\frac{4}{L}\sum_{m=-n_{F}(V_{G})}^{m= n_{F}(V_{G})} f_{F.D.}[\frac{\epsilon(m)-\delta\mu_{0}(T)}{K_{B}T}]$ . The effect of the interactions will replace the ground state $|F>$ by the renormalized ground state $|G>$. The ground state represents a shifted Fermi Surface given by the self energy $\delta\Sigma (V_{G},l(V_{G}))$ determined by the self consistent equation:
\begin{equation}
\delta\Sigma (V_{G},l(V_{G}))= 2\eta_{c}(l(V_{G}))<G|\hat{Q}_{c}|G>\equiv2\eta_{c}(l(V_{G}))4\sum_{m=-n_{F}(V_{G})}^{m= n_{F}(V_{G})} f_{F.D.}[\frac{\epsilon(m)+\delta\Sigma (V_{G},l(V_{G}))-\delta\mu_{0}(T)}{K_{B}T}]
\label{self}
\end{equation}
The solution  for $\delta\Sigma (V_{G},l(V_{G}))$  is obtained  once we replace the sum  by an energy integration (the density of states cancel the velocity):
\begin{equation} 
\delta\Sigma (V_{G},l(V_{G}))\equiv h v_{F}[(\frac{1-K^{2}_{R}(l(V_{G})}{K^{2}_{R}l(V_{G})})+\gamma (\frac{c}{v_{F}})F(\frac{L}{d e^{l(V_{G})}},\frac{\xi}{d e^{l(V_{G})}})]\cdot \frac{n_{e}(V_{G})}{\hat{\epsilon}(V_{G},l,T,L)}
\label{delta}
\end{equation}
where the explicit form  ${\hat{\epsilon}(V_{G},l,T,L)}$  represents  the effective dielectric function given by:
\begin{eqnarray}
\hat{\epsilon}(V_{G},l,T,L)&=&1+4[\frac{1-K^{2}_{R}(l (V_{G}))}{K^{2}_{R}(l(V_{G}))}+\gamma (\frac{c}{v_{F}})F(\frac{L}{d e^{l(V{G})}},\frac{\xi}{d e^{l(V{G})}})\cdot r(T)]
\label{eqnarray}
\end{eqnarray}
where $r(T)$ represents the thermal correction, which is 1   when we use the approximation :      $\int_{\epsilon_{F}(V_{G})-\delta\Sigma (V_{G},l(V_{G}))}^{\epsilon_{F}(V_{G})+\delta\Sigma (V_{G},l(V_{G}))} d \epsilon f_{F.D.}[\frac{\epsilon-\delta\mu_{0}(T)}{K_{B}T}]\approx 0$. When the self energy is small   with respect the Fermi energy  $\frac{\delta\Sigma (V_{G},l(V_{G}))}{\epsilon_{F}(V_{G})}<<1$,  
  we expand the Fermi Dirac function with  respect   $\delta\Sigma (V_{G},l(V_{G}))$  and find: $r(T)=f_{F.D.}[\frac{- \epsilon_{F}(V_{G}) -\delta\mu_{0}(T)}{K_{B}T}]-f_{F.D.}[\frac{ \epsilon_{F}(V_{G}) -\delta\mu_{0}(T)}{K_{B}T}])]$.

\end{document}